\RequirePackage[2020-02-02]{latexrelease}

\documentclass[twocolumn,preprintnumbers]{revtex4}%
\usepackage{amssymb}
\usepackage{amsmath}
\usepackage{graphicx}
\usepackage{dcolumn}
\usepackage{bm}
\usepackage{xcolor}
\usepackage{ifthen}
\usepackage[normalem]{ulem}
\usepackage{amsfonts}%
\RequirePackage[2020-02-02]{latexrelease}
\graphicspath{{C:/Users/Eyal/MyDocuments/latex/Texfiles/Diamond/dipolarBA/EntangSupp/SD/}}
\RequirePackage[2020-02-02]{latexrelease}
\RequirePackage[2020-02-02]{latexrelease}
\UseRawInputEncoding
\providecommand{\U}[1]{\protect\rule{.1in}{.1in}}
\providecommand{\U}[1]{\protect\rule{.1in}{.1in}}
\def\showal{1}
\newcommand{\al}[1]{\ifthenelse{\showal=1}{\textcolor{orange}{[[#1]]}}{}}

\newcommand{\eb}[1]{\ifthenelse{\showal=1}{\textcolor{cyan}{[[#1]]}}{}}
\begin{document}
\title{Disentanglement-induced multistability}
\author{Eyal Buks}
\email{eyal@ee.technion.ac.il}
\affiliation{Andrew and Erna Viterbi Department of Electrical Engineering, Technion, Haifa
32000, Israel}
\date{\today }

\begin{abstract}
Multistability cannot be derived from any theoretical model that is based on a
monostable master equation. On the other hand, multistability is
experimentally-observed in a variety of quantum systems. A master equation
having a nonlinear term that gives rise to disentanglement has been recently
proposed . The dynamics governed by this master equation is explored for a
quantum system made of coupled spins. It is found that the added nonlinear
term can give rise to multistability. The spins' response to an externally
applied magnetic field is evaluated, and both a phase transition and a
dynamical instability are found. These findings, which originate from
disentanglement-induced multistability, indirectly support the hypothesis that
spontaneous disentanglement occurs in quantum systems.

\end{abstract}
\maketitle

\textbf{Introduction} - The time evolution of a quantum system in contact with
its environment is commonly described using a master equation (ME) for the
system's reduced density operator $\rho$. A ME having a unique steady state
solution is said to be monostable. The widely employed
Gorini-Kossakowski-Sudarshan-Lindblad (GKSL) ME
\cite{Fernengel_385701,Lindblad_119,Manzano_025106}, which is linear in $\rho
$, is monostable, provided that the Hamiltonian is time independent. For such
Hamiltonians, the Grabert ME, which has a nonlinear dependency on $\rho$, is
also monostable (its unique steady state solution is thermal equilibrium)
\cite{Grabert_161,Ottinger_052119}.

In contrast, some experimentally observed behaviors in quantum systems suggest
multistability in the underlying dynamics. For example, consider a
single-domain ferromagnet under the influence of a transverse (with respect to
the domain's easy axis) static magnetic filed. Above a critical temperature
$T_{\mathrm{c}}$, the system is monostable. However, a phase transition (PT)
occurs at $T_{\mathrm{c}}$, below which, the magnetization has two
locally-stable steady states. Both the PT and the multistability cannot be
derived from any monostable ME.

More generally, for any finite system having a static (i.e. time independent)
Hamiltonian, both PTs and multistabilities cannot be derived from any
monostable ME
\cite{Chomaz_68,mainwood2005phase,Callender_539,Liu_S92,Ardourel_99,Shech_1170}%
. The term finite commonly refers to one out of two system's properties. The
first property is number of particles, and the second one is Hilbert space
dimensionality. Exclusion of PTs holds for both definitions of this term
\cite{toda1978statistical_I}. On the other hand, all experimental observations
of PTs are performed using finite systems. Moreover, a PT has been
experimentally observed in small systems, including molecular magnets
\cite{Roch_633,Thomas_145,Trishin_236801,Blesio_045113}.

For some cases, the time evolution of a given dynamical system (i.e. a system
having time dependent Hamiltonian) can be described using a static (i.e. time
independent) Hamiltonian. For example, when the rotating wave approximation
(RWA) is applicable, a transformation into a rotating frame yields a static
Hamiltonian. Similarly to the case of static systems, multistability can be
theoretically excluded, provided that the system is finite and the RWA is
applicable (see appendix B of Ref. \cite{Levi_053516}). In contrast,
multistability has been experimentally observed in dynamical spin systems (for
which both number of particles and Hilbert space dimensionality are finite)
\cite{gurevich2020magnetization}. One example is a dynamical instability (DI)
induced by parallel pumping applied to a ferrimagnetic insulator containing
finite (though commonly large) number of spins
\cite{Suhl_209,Schlomann_672,Zvyagin_174408}. In these experiments, the
parallel pumping angular frequency is tuned close to $2\omega_{\mathrm{L}}$,
where $\omega_{\mathrm{L}}$ is the spins' resonance angular frequency. For a
driving amplitude $\omega_{1}$ smaller than a critical value $\omega
_{1,\mathrm{c}}$ (instability threshold), the system's response is monostable.
However, at $\omega_{1,\mathrm{c}}$ a bifurcation occurs, above which the
spins oscillate with a relative phase $\phi$, which is either $\phi=0$ or
$\phi=\pi$.

Theoretical models, which have been developed to account for experimentally
observed multistability in finite quantum systems, are usually based on the
assumption that the underlying dynamics is nonlinear. Commonly, such
nonlinearity is introduced by implementing the mean field approximation (MFA).
It has been shown that the MFA yields both PTs \cite{toda1978statistical_I}
and DIs
\cite{Schlomann_S386,breuer2002theory,Drossel_217,Hicke_024401,Klobus_034201,Hush_061401}
in finite systems. The MFA is based on the assumption that entanglement
between subsystems can be disregarded. However, it has remained unclear how
such an assumption can be justified within the framework of standard quantum
mechanics (QM) \cite{Vedral_22,Sakthivadivel_035,Osacar_045404}, particularly
for cases where the MFA turns a given monostable time evolution into a
multistable one.

A modified ME has been recently proposed \cite{Buks_2400036}. This ME [see Eq.
(\ref{MME}) below] has an added nonlinear term
\cite{Kaplan_055002,Geller_2200156} given by $-\Theta\rho-\rho\Theta
+2\left\langle \Theta\right\rangle \rho$, which can give rise to both
disentanglement and thermalization. Under some appropriate conditions, the ME
(\ref{MME}) becomes multistable. The dynamics generated by the modified ME
(\ref{MME}) is explored below for a system made of a finite number of coupled
spins. It is found that the interplay between an externally applied static
magnetic field and the dipolar coupling between spins gives rise to a PT,
which separates a region of monostability and a region of bistability.
Moreover, when an externally applied parametric excitation (parallel pumping)
is added, the modified ME (\ref{MME}) yields a DI, above which a periodic
limit cycle occurs.

\textbf{Nonlinear ME} - The proposed modified ME is given by
\cite{Grimaudo_033835,Kowalski_167955,Buks_2400036}%
\begin{equation}
\frac{\mathrm{d}\rho}{\mathrm{d}t}=i\hbar^{-1}\left[  \rho,\mathcal{H}\right]
-\Theta\rho-\rho\Theta+2\left\langle \Theta\right\rangle \rho\;, \label{MME}%
\end{equation}
where $\hbar$ is the Planck's constant, $\mathcal{H}^{{}}=\mathcal{H}^{\dag}$
is the Hamiltonian, the operator $\Theta^{{}}=\Theta^{\dag}$ is allowed to
depend on $\rho$, and $\left\langle \Theta\right\rangle =\operatorname{Tr}%
\left(  \Theta\rho\right)  $. For the case $\mathcal{H}=0$, and for a fixed
$\Theta$, the modified master equation (\ref{MME}) yields an equation of
motion for $\left\langle \Theta\right\rangle $ given by $\mathrm{d}%
\left\langle \Theta\right\rangle /\mathrm{d}t=-2\left\langle \left(
\Theta-\left\langle \Theta\right\rangle \right)  ^{2}\right\rangle $, which
implies that the expectation value $\left\langle \Theta\right\rangle $
monotonically decreases with time. Hence, the nonlinear term in the modified
ME (\ref{MME}) can be employed to suppress a given physical property, provided
that $\left\langle \Theta\right\rangle $ quantifies that property. The
operator $\Theta$ is assumed to be given by $\Theta=\gamma_{\mathrm{H}%
}\mathcal{Q}^{\left(  \mathrm{H}\right)  }+\gamma_{\mathrm{D}}\mathcal{Q}%
^{\left(  \mathrm{D}\right)  }$, where both rates $\gamma_{\mathrm{H}}$ and
$\gamma_{\mathrm{D}}$ are positive, and both operators $\mathcal{Q}^{\left(
\mathrm{H}\right)  }$ and $\mathcal{Q}^{\left(  \mathrm{D}\right)  }$\ are
Hermitian. The first term $\gamma_{\mathrm{H}}\mathcal{Q}^{\left(
\mathrm{H}\right)  }$ gives rise to thermalization
\cite{Grabert_161,Ottinger_052119}, whereas the second one $\gamma
_{\mathrm{D}}\mathcal{Q}^{\left(  \mathrm{D}\right)  }$ gives rise to disentanglement.

The thermalization operator is given by $\mathcal{Q}^{\left(  \mathrm{H}%
\right)  }=\beta\mathcal{U}_{\mathrm{H}}$, where $\mathcal{U}_{\mathrm{H}%
}=\mathcal{H}+\beta^{-1}\log\rho$ is the Helmholtz free energy operator,
$\beta=1/\left(  k_{\mathrm{B}}T\right)  $ is the thermal energy inverse,
$k_{\mathrm{B}}$ is the Boltzmann's constant, and $T$ is the temperature. The
thermal equilibrium density operator $\rho_{0}=e^{-\beta\mathcal{H}%
}/\operatorname{Tr}\left(  e^{-\beta\mathcal{H}}\right)  $, which minimizes
the Helmholtz free energy $\left\langle \mathcal{U}_{\mathrm{H}}\right\rangle
$, is a steady state solution of the ME (\ref{MME}), provided that
$\gamma_{\mathrm{D}}=0$ (i.e. no disentanglement) and the Hamiltonian
$\mathcal{H}$ is time independent
\cite{Jaynes_579,Grabert_161,Ottinger_052119,Buks_052217}. The construction of
the disentanglement operator $\mathcal{Q}^{\left(  \mathrm{D}\right)  }$ is
explained in Ref. \cite{Buks_2400036}.

\begin{figure}[ptb]
\begin{center}
\includegraphics[width=2.5in,keepaspectratio]{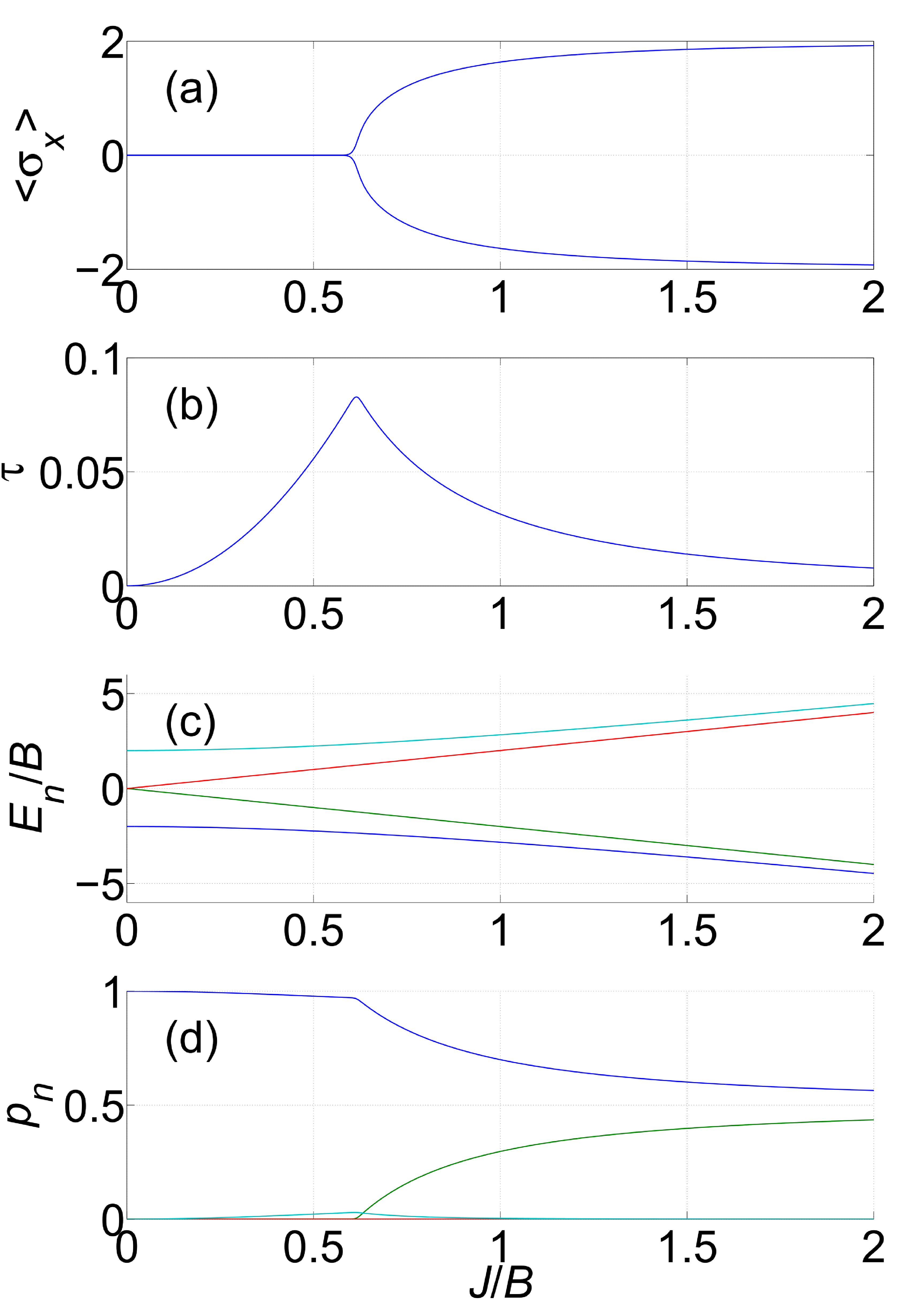}
\end{center}
\caption{{}Two spin PT. The (a) magnetization $\left\langle \sigma
_{x}\right\rangle $, (b) two-spin entanglement $\tau$, (c) energy eigenvalues
$E_{n}$, and (d) population probabilities $p_{n}$, are plotted as a function
of the ratio $J/B$. Assumed parameters' values are $\hbar\gamma_{\mathrm{H}%
}/B=50$ and $\hbar\gamma_{\mathrm{D}}\beta^{-1}/B^{2}=100$.}%
\label{FigTSPT}%
\end{figure}

\textbf{Spin PT} - The one-dimensional transverse Ising model (TIM)
Hamiltonian is given by
\cite{Lieb_407,Pfeuty_79,Mbeng_2009_09208,Nagaj_214431,chakrabarti2008quantum}%
\begin{equation}
\mathcal{H}=-B\sum_{l=1}^{L}\sigma_{l,z}-J\sum_{l=1}^{L}\sigma_{l,x}%
\sigma_{l+1,x}\ , \label{H JWT}%
\end{equation}
where both $B$ and $J$ are real non-negative constants. The number of spins,
which is assumed to be finite, is denoted by $L$, and the Pauli vector
operator $\left(  \sigma_{l,x},\sigma_{l,y},\sigma_{l,z}\right)  $ represents
the $l$'th spin angular momentum in units of $\hbar/2$, where $l\in\left\{
1,2,\cdots,L\right\}  $. It is assumed that the one-dimensional spin array has
a ring configuration, and thus, the last ($l=L$) coupling term $\sigma
_{l,x}\sigma_{l+1,x}$ [see Eq. (\ref{H JWT})] is taken to be given by
$\sigma_{L,x}\sigma_{1,x}$.

For the two spin case (i.e. $L=2$), the matrix representation of $\mathcal{H}$
in the basis $\left\{  \left\vert -1,-1\right\rangle ,\left\vert
-1,1\right\rangle ,\left\vert 1,-1\right\rangle ,\left\vert 1,1\right\rangle
\right\}  $ is given by%
\begin{equation}
\mathcal{H}\dot{=}-2\left(
\begin{array}
[c]{cccc}%
B & 0 & 0 & J\\
0 & 0 & J & 0\\
0 & J & 0 & 0\\
J & 0 & 0 & -B
\end{array}
\right)  \ , \label{H TS}%
\end{equation}
where the ket vector $\left\vert \sigma_{2},\sigma_{1}\right\rangle $ is an
eigenvector of $\sigma_{l,z}$ with an eigenvalue $\sigma_{l}\in\left\{
-1,1\right\}  $, and where $l\in\left\{  1,2\right\}  $. Steady state
solutions of the modified ME (\ref{MME}) are shown for this case in Fig.
\ref{FigTSPT} as a function of the ratio $J/B$. The magnetization
$\left\langle \sigma_{x}\right\rangle =\left\langle \sigma_{1,x}\right\rangle
+\left\langle \sigma_{2,x}\right\rangle $, which is plotted in Fig.
\ref{FigTSPT}(a), becomes finite above a critical value of the ratio $J/B$
(which depends on the rates $\gamma_{\mathrm{H}}$ and $\gamma_{\mathrm{D}}$
and on the temperature). The plot in Fig. \ref{FigTSPT}(b) depicts the
two-spin entanglement
\cite{Grimmett_305,Jian_134206,Latorre_0304098,Osborne_032110,Parez_2402_06677}
$\tau=\left\langle \mathcal{Q}^{\left(  \mathrm{D}\right)  }\right\rangle $
[see Eq. (11) of Ref. \cite{Buks_2400036}]. Note that $\tau$ peaks near the PT.

Standard QM predicts that $\left\langle \sigma_{x}\right\rangle =0$ in steady
state [note that the Hamiltonian $\mathcal{H}$ (\ref{H JWT}) is invariant
under the mirror reflection $x\rightarrow-x$, and consequently, $\left\langle
n\right\vert \sigma_{x}\left\vert n\right\rangle =0$ for all energy
eigenvectors $\left\vert n\right\rangle $]. In contrast, non-vanishing values
for $\left\langle \sigma_{x}\right\rangle $ in steady state become possible in
the presence of spontaneous disentanglement [see Fig. \ref{FigTSPT}(a)]. For
the TIM, the assumption that the MFA is applicable leads to magnetization
$\left\langle \sigma_{x}\right\rangle _{\mathrm{MFA}}$ given by
\cite{Mbeng_2009_09208}%
\begin{equation}
\left\langle \sigma_{x}\right\rangle _{\mathrm{MFA}}=\left\{
\begin{array}
[c]{cc}%
0 & \text{for }\frac{2J}{B}<1\\
\pm\sqrt{1-\left(  \frac{B}{2J}\right)  ^{2}} & \text{for }\frac{2J}{B}\geq1
\end{array}
\right.  \;. \label{<sigma_x>_m}%
\end{equation}
However, as has been discussed above, it has remained unclear how the MFA can
be justified within the framework of standard QM. Even though both MFA and
spontaneous disentanglement can account for multistability, their predictions
are distinguishable, as is discussed below.

\begin{figure}[ptb]
\begin{center}
\includegraphics[width=3.2in,keepaspectratio]{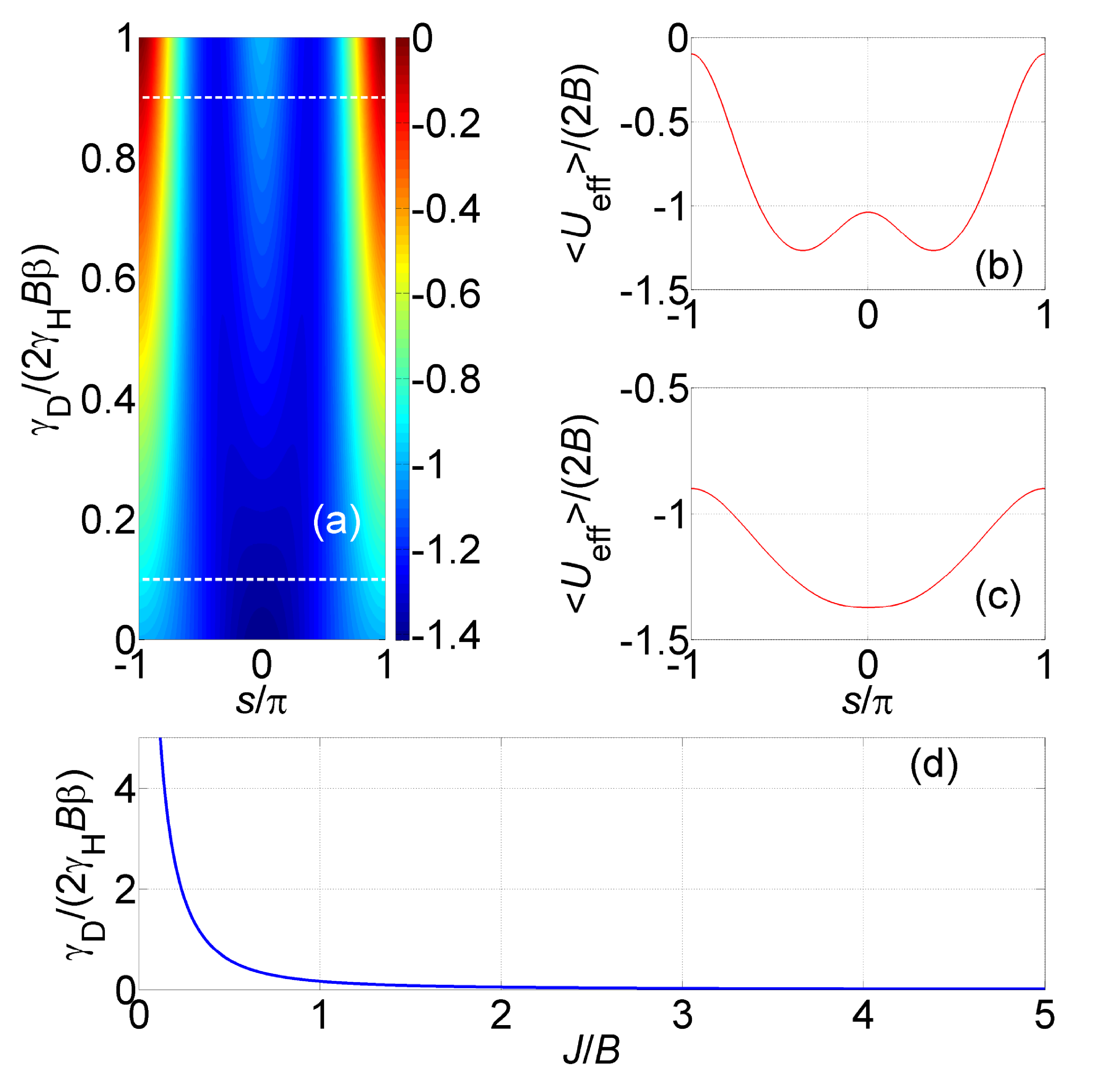}
\end{center}
\caption{{}The effective free energy $\left\langle \mathcal{U}_{\mathrm{eff}
}\right\rangle $. (a) Dependency of $\left\langle \mathcal{U} _{\mathrm{eff}%
}\right\rangle $ on $\gamma_{\mathrm{D}}/\left(  2\gamma_{\mathrm{H}}%
B\beta\right)  $ and $s$. The cross section plots (c) and (b) demonstrate the
regions of monostability and bistability, respectively. The overlaid white
dotted lines in (a) indicate the two values of the ratio $\gamma_{\mathrm{D}%
}/\left(  2\gamma_{\mathrm{H}}B\beta\right)  $ corresponding to the cross
section plots in (b) and (c). For (a), (b) and (c) it is assumed that $J/B=1$.
(d) The critical value of the ratio $\gamma_{\mathrm{D}}/\left(
2\gamma_{\mathrm{H}}B\beta\right)  $ as a function of $J/B$.}%
\label{FigEffH}%
\end{figure}

\begin{figure*}[ptb]
\par
\begin{center}
\includegraphics[width=6.4in,keepaspectratio]{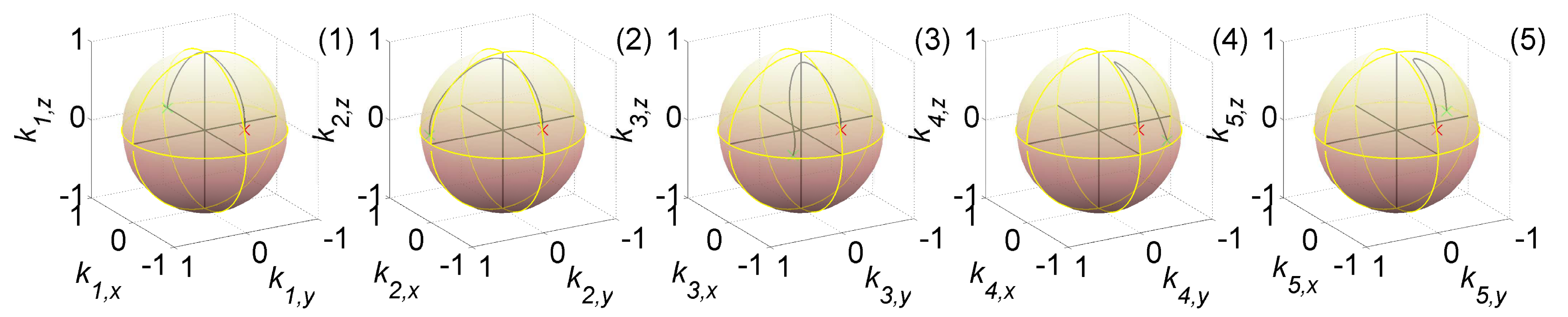}
\end{center}
\caption{{}Five spin TIM. The green (red) cross symbols represent initial
(final) values of the single-spin Bloch vectors. Initially the spins are in a
product state, where the $l$'th spin is pointing in the direction $\left(
\cos\left(  2\pi\left(  l-1\right)  /L\right)  ,\sin\left(  2\pi\left(
l-1\right)  /L\right)  ,0\right)  $. Parameters' assumed values are $J/B=2$,
$\hbar\gamma_{\mathrm{H}}/B=5$ and $\hbar\gamma_{\mathrm{D}} \beta^{-1}%
/B^{2}=100$.}%
\label{FigFSBS}%
\end{figure*}

\begin{figure}[ptb]
\begin{center}
\includegraphics[width=3.2in,keepaspectratio]{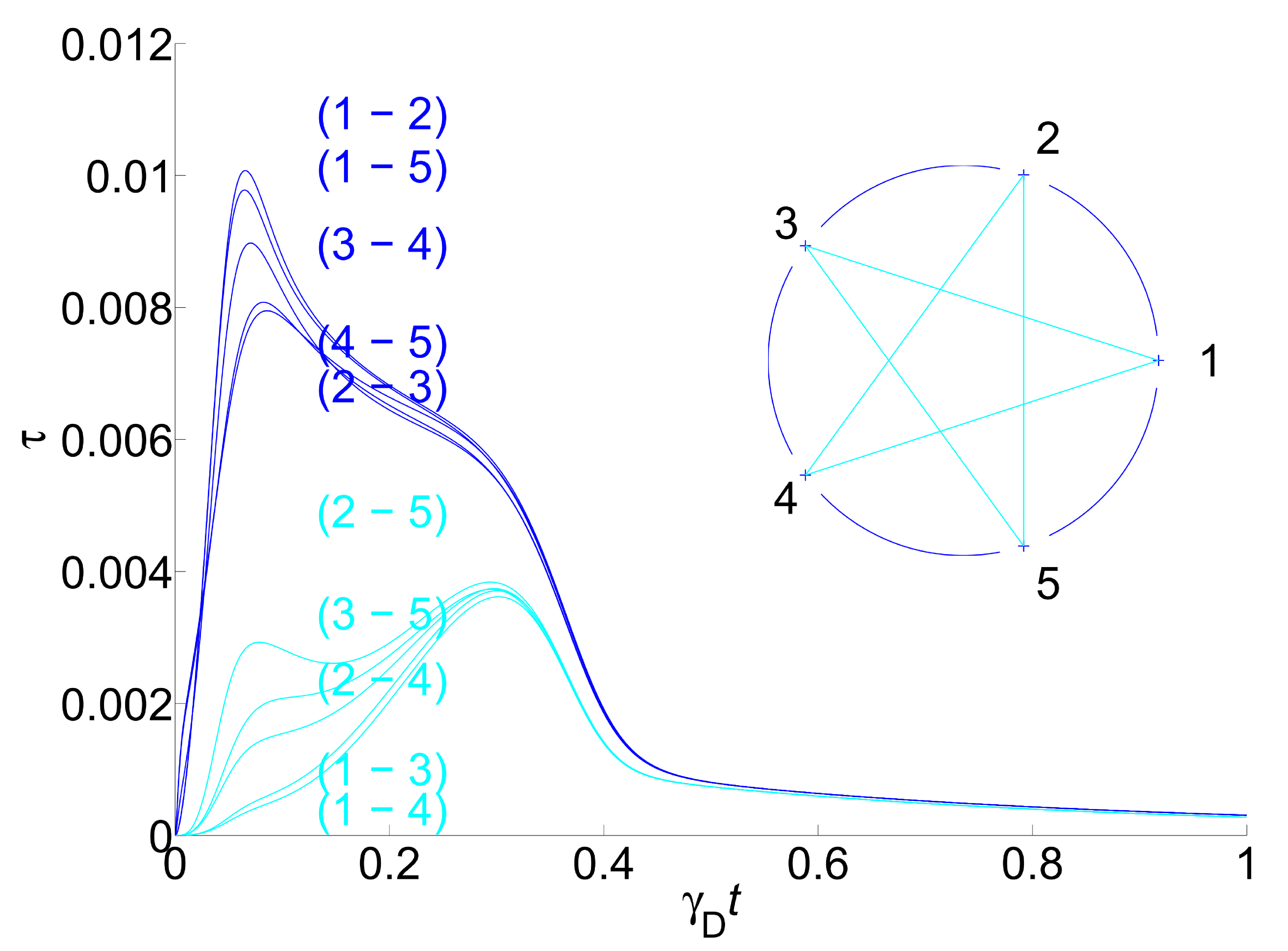}
\end{center}
\caption{{}Pair entanglement $\tau$. Plots corresponding to nearest (second
nearest) neighbor pairs are blue (cyan) colored. Parameters' assumed values
are listed in the caption of Fig \ref{FigFSBS}. The number of both nearest and
second nearest neighbor pairs is $L=5$. The initial product state at
$\gamma_{\mathrm{D}}t=0$, for which entanglement of all pairs vanish, is
associated with the 5 green cross symbols shown in Fig. \ref{FigFSBS}. At is
shown in Fig. \ref{FigFSBS}, initially all 5 single-spin Bloch vectors move
from their initial values along the equator towards the north pole of their
Bloch spheres. This process corresponds to the time interval of $0<\gamma
_{\mathrm{D}}t\lesssim0.3$. The limit $\gamma_{\mathrm{D}}t\rightarrow\infty$
is represented by the 5 red crosses in Fig. \ref{FigFSBS}.}%
\label{FigFStau}%
\end{figure}

The relation $\Theta=\gamma_{\mathrm{H}}\mathcal{Q}^{\left(  \mathrm{H}%
\right)  }+\gamma_{\mathrm{D}}\mathcal{Q}^{\left(  \mathrm{D}\right)  }$
together with the ME (\ref{MME}) suggest that disentanglement can be accounted
for by replacing the Helmholtz free energy $\left\langle \mathcal{U}%
_{\mathrm{H}}\right\rangle $ by an effective free energy $\left\langle
\mathcal{U}_{\mathrm{eff}}\right\rangle $, which is given by $\left\langle
\mathcal{U}_{\mathrm{eff}}\right\rangle =\left\langle \mathcal{U}_{\mathrm{H}%
}\right\rangle +\beta^{-1}\left(  \gamma_{\mathrm{D}}/\gamma_{\mathrm{H}%
}\right)  \left\langle \mathcal{Q}^{\left(  \mathrm{D}\right)  }\right\rangle
$. Consider the case where the temperature is sufficiently low to validate the
approximation $\left\langle \mathcal{U}_{\mathrm{H}}\right\rangle
\simeq\left\langle \mathcal{H}\right\rangle $. The energy eigenvectors of the
two spin Hamiltonian (\ref{H TS}) are denoted by $\left\vert 1\right\rangle $,
$\left\vert 2\right\rangle $, $\left\vert 3\right\rangle $ and $\left\vert
4\right\rangle $, and the corresponding eigenenergies by $E_{1}=-2\sqrt
{B^{2}+J^{2}}$, $E_{2}=-2J$, $E_{3}=2J$ and $E_{4}=2\sqrt{B^{2}+J^{2}}$ [see,
respectively, blue, green, red and cyan lines in Fig. \ref{FigTSPT}(c)]. As
can be seen from the red and cyan lines in Fig. \ref{FigTSPT}(d), the
probabilities $p_{3}$ and $p_{4}$ to occupy the energy eigenstates $\left\vert
3\right\rangle $ and $\left\vert 4\right\rangle $ are relatively small both
below and above the PT. Consider a class of pure states, for which only the
two lowest energy states are occupied. A state $\left\vert \psi\right\rangle $
belonging to this class is expressed as $\left\vert \psi\right\rangle
=e^{i\varphi/2}\sqrt{\left(  1+\cos s\right)  /2}\left\vert 1\right\rangle
+e^{-i\varphi/2}\sqrt{\left(  1-\cos s\right)  /2}\left\vert 2\right\rangle $,
where both $\varphi$ and $s$ are real. With the help of Eq. (15) of Ref.
\cite{Buks_2400036}, the effective free energy $\left\langle \mathcal{U}%
_{\mathrm{eff}}\right\rangle $ can be analytically calculated for this class
of pure states. Note that for a given $s$, the entanglement $\left\langle
\psi\right\vert \mathcal{Q}^{\left(  \mathrm{D}\right)  }\left\vert
\psi\right\rangle $ is minimized for $\varphi=0$, and that the energy
expectation value $\left\langle \psi\right\vert \mathcal{H}\left\vert
\psi\right\rangle $ does not depend on the phase $\varphi$. Thus, for the
current case, the minimization of the effective free energy $\left\langle
\mathcal{U}_{\mathrm{eff}}\right\rangle $ can be simplified by setting
$\varphi=0$. Note that the magnetization $\left\langle \sigma_{x}\right\rangle
$ for this setting is given by $\left\langle \sigma_{x}\right\rangle =2\left(
1+e^{-2\sinh^{-1}\left(  J/B\right)  }\right)  ^{-1/2}\sin s$. The color-coded
plot in Fig. \ref{FigEffH}(a) depicts the dependency of the effective free
energy $\left\langle \mathcal{U}_{\mathrm{eff}}\right\rangle $ on the ratio
$\gamma_{\mathrm{D}}/\left(  2\gamma_{\mathrm{H}}B\beta\right)  $ and on the
parameter $s$ (for $\varphi=0$). The plot reveals a PT from monostability to
bistability occurring at a critical value of the ratio $\gamma_{\mathrm{D}%
}/\left(  2\gamma_{\mathrm{H}}B\beta\right)  $. The region of monostability is
demonstrated by the plot in Fig. \ref{FigEffH}(c), whereas bistability is
demonstrated by the plot in Fig. \ref{FigEffH}(b). The two values of the ratio
$\gamma_{\mathrm{D}}/\left(  2\gamma_{\mathrm{H}}B\beta\right)  $
corresponding to the plots in Fig. \ref{FigEffH}(b) and (c) are indicated by
the overlaid white dotted lines in Fig. \ref{FigEffH}(a).

Stability analysis is employed to extract the critical value of the ratio
$\gamma_{\mathrm{D}}/\left(  2\gamma_{\mathrm{H}}B\beta\right)  $ from the
effective free energy $\left\langle \mathcal{U}_{\mathrm{eff}}\right\rangle $.
The plot shown in Fig. \ref{FigEffH}(d) depicts the ratio $\gamma_{\mathrm{D}%
}/\left(  2\gamma_{\mathrm{H}}B\beta\right)  $ as a function of $J/B$ at the
PT. While the MFA yields a constant value for $J/B$ at the PT [see Eq.
(\ref{<sigma_x>_m})], disentanglement makes this value becoming dependent on
the rate $\gamma_{\mathrm{D}}$ [see Fig. \ref{FigEffH}(d)]. Moreover, in the
region where $\left\langle \sigma_{x}\right\rangle \neq0$, the magnetization
$\left\langle \sigma_{x}\right\rangle $ dependency on the ratio $J/B$
according to the spontaneous disentanglement hypothesis [see Fig.
\ref{FigTSPT}(a)] is distinguishable from the one that is derived from the MFA
[see Eq. (\ref{<sigma_x>_m})].

Only nearest neighbor spins are coupled in the TIM [see Eq. (\ref{H JWT})].
The disentanglement operator $\mathcal{Q}^{\left(  \mathrm{D}\right)  }$ for
the case $L>2$ (i.e. for more than two spins) is accordingly constructed, by
including only nearest neighbor terms. The case $L=5$ is demonstrated by the
plots shown in Fig. \ref{FigFSBS} and Fig. \ref{FigFStau}. Time evolution of
the single-spin Bloch vector $\mathbf{k}_{l}=\left(  \left\langle \sigma
_{l,x}\right\rangle ,\left\langle \sigma_{l,y}\right\rangle ,\left\langle
\sigma_{l,z}\right\rangle \right)  $ is shown in Fig. \ref{FigFSBS}($l$),
where $l\in\left\{  1,2,3,4,5\right\}  $. These plots demonstrate flow towards
a locally stable steady state with positive magnetization $\left\langle
\sigma_{x}\right\rangle $. A state with the opposite magnetization value is
also locally stable. The entire Hilbert space is divided into two basins of
attractions associated with these two locally stable steady states.

For the case $L=5$, each spin has two nearest neighbors, and two second
nearest neighbors (see the inset of Fig. \ref{FigFStau}). The time evolution
of the entanglement variable $\tau$ for all spin pairs is shown in the plot in
Fig. \ref{FigFStau}. Plots corresponding to nearest (second nearest) neighbor
pairs are blue (cyan) colored. The shared steady state value of $\tau$ for all
nearest neighbor pairs, which is denoted by $\tau_{\mathrm{NN}}$, is found to
be larger than the shared steady state value of $\tau$ for all second nearest
neighbor pairs, which is denoted by $\tau_{\mathrm{SNN}}$ ($\tau_{\mathrm{NN}%
}/\tau_{\mathrm{SNN}}\simeq1.11$ for the example shown in Fig. \ref{FigFStau}).

\textbf{Spin parametric instability} - Disentanglement-induced multistability
is explored below for a system made of two spins under parallel pumping. The
time-dependent Hamiltonian $\mathcal{H}$ is given by%
\begin{equation}
\frac{\mathcal{H}}{\hbar}=\frac{\omega_{z}\sigma_{z}}{2}+\omega_{\mathrm{L}%
}\vartheta\frac{\sigma_{y}^{2}-\sigma_{x}^{2}}{4}\;, \label{H_0 PE}%
\end{equation}
where $\omega_{z}=-\omega_{\mathrm{L}}+\omega_{1}\cos\left(  2\omega
_{\mathrm{L}}t\right)  $. The Larmor angular frequency $\omega_{\mathrm{L}}$,
the longitudinal driving amplitude$\ \omega_{1}$, and the (assumed small)
demagnetization asymmetry factor$\ \vartheta$ \cite{kittel1976introduction}
are all real constants, and $\sigma_{i}=\sigma_{1,i}+\sigma_{2,i}$ for
$i\in\left\{  x,y,z\right\}  $. Note that the term in Eq. (\ref{H_0 PE})
proportional to $\sigma_{y}^{2}-\sigma_{x}^{2}$ generates squeezing
\cite{Kitzinger_033504,Kitagawa_5138,Ma_89}. In the RWA, the time dependent
lab frame Hamiltonian $\mathcal{H}$ is transformed into a time independent
rotating frame Hamiltonian $\mathcal{H}_{\mathrm{RWA}}$, which has matrix
representation given by [see Eq. (17.212) of Ref. \cite{Buks_QMLN}, and
compare to Eq. (\ref{H TS})]%
\begin{equation}
\frac{\mathcal{H}_{\mathrm{RWA}}}{\hbar}\dot{=}-2\left(
\begin{array}
[c]{cccc}%
\mathcal{B} & 0 & 0 & \mathcal{J}\\
0 & 0 & 0 & 0\\
0 & 0 & 0 & 0\\
\mathcal{J} & 0 & 0 & -\mathcal{B}%
\end{array}
\right)  +O\left(  \vartheta^{2}\right)  \;, \label{H_RWA}%
\end{equation}
where $\mathcal{B}=-\omega_{1}/2$ and $\mathcal{J}=-\omega_{1}\vartheta/4$.

As can be seen from Eq. (\ref{H_RWA}), $\mathcal{H}_{\mathrm{RWA}}$ becomes
diagonal when the demagnetization asymmetry factor $\vartheta$ vanishes. This
behavior is attributed to the observation that for $\vartheta=0$ there is no
preferred direction in the plane perpendicular to the constant magnetic field
(the $xy$ plane). Thus, for $\vartheta=0$ the relative phase $\phi$ of
precession with respect to the parametric driving has no preferred value. On
the other hand, for finite $\vartheta$, the relative phase $\phi$ has two
preferred values denoted by $\phi_{1}=0$ and $\phi_{2}=\pi$.

\begin{figure}[ptb]
\begin{center}
\includegraphics[width=3.2in,keepaspectratio]{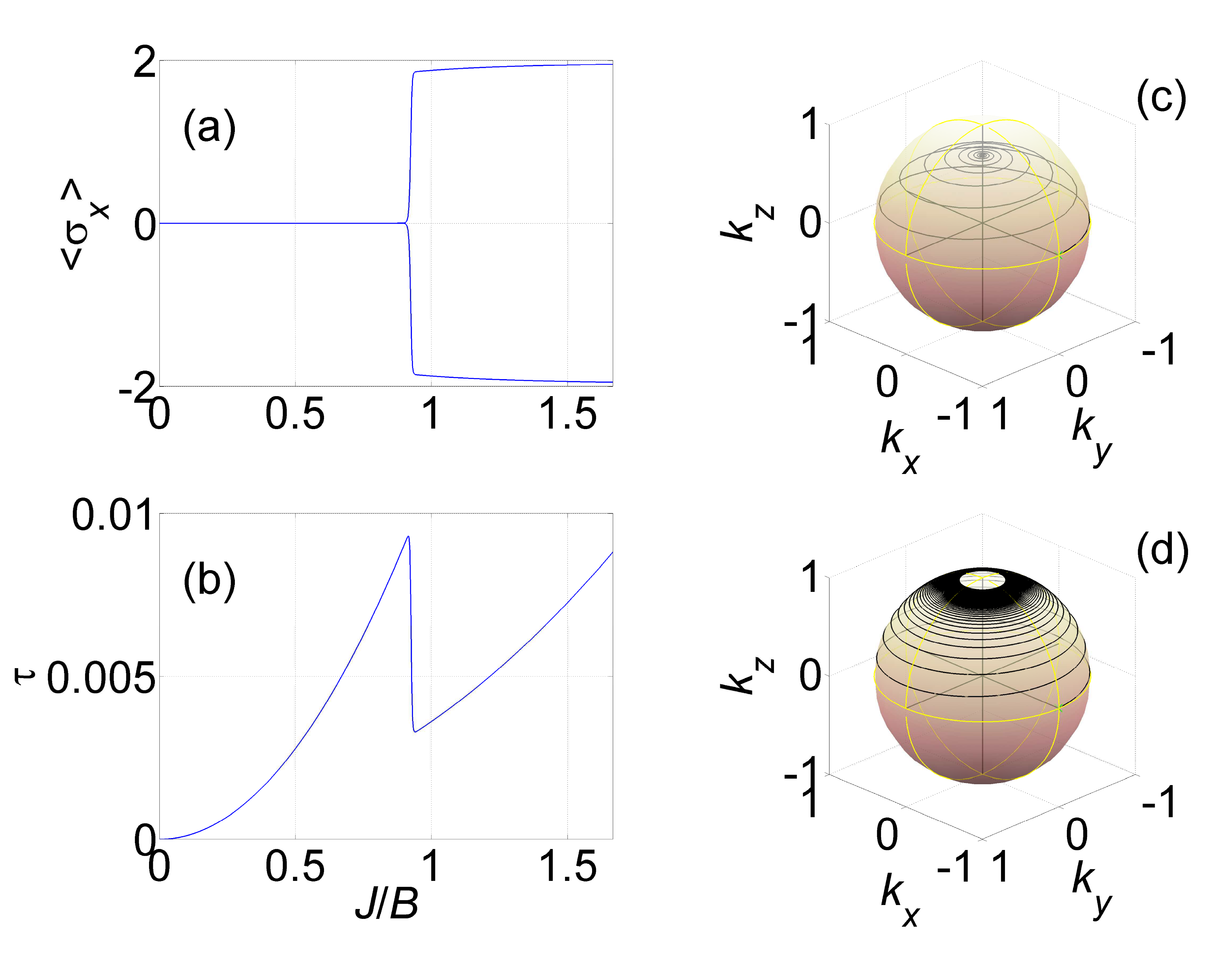}
\end{center}
\caption{{}Parallel pumping. The (a) magnetization $\left\langle \sigma_{x}\right\rangle
	$ and (b) two-spin entanglement $\tau$ in steady state are plotted as a function of the ratio
	$\mathcal{J}/\mathcal{B}$.  Assumed parameters' values are $\hbar
	\gamma_{\mathrm{H}}/\mathcal{B=}5$ and $\hbar\gamma_{\mathrm{D}}\beta
	^{-1}/\mathcal{B}^{2}=100$. The green cross symbols in (c) and (d) represent
	initial values of the Bloch vector. While disentanglement is inactive (i.e.
	$\gamma_{\mathrm{D}}=0$) in (c), the dimensionless disentanglement rate is
	$\hbar\gamma_{\mathrm{D}}\beta^{-1}/\mathcal{B}^{2}=100$ in (d). For both
	(c) and (d) $\mathcal{J}/\mathcal{B}=1.05$.}%
\label{FigTSP}%
\end{figure}

The time evolution of the Bloch vector $\mathbf{k}=\left(  \left\langle
\sigma_{x}\right\rangle ,\left\langle \sigma_{y}\right\rangle ,\left\langle
\sigma_{z}\right\rangle \right)  $ is shown in Fig. \ref{FigTSP}(c) and (d). As is
demonstrated by the plot in Fig. \ref{FigTSP}(c), in the absence of
disentanglement (i.e. $\gamma_{\mathrm{D}}=0$), the steady state is a fixed
point. By turning on disentanglement (all other parameters are kept
unchanged), the steady state becomes a periodic limit cycle, as is
demonstrated by the plot in Fig. \ref{FigTSP}(d). The dependency on the ratio $\mathcal{J}/\mathcal{B}$ of the magnetization $\left\langle \sigma_{x}\right\rangle$ and the two-spin entanglement $\tau$ is shown in Fig. \ref{FigTSP} (a) and (b), respectively. Due to the similarity
between the Hamiltonians (\ref{H TS}) and (\ref{H_RWA}), the underlying
mechanism responsible for the instability seen in Fig. \ref{FigTSP}, is
similar to the one seen in Fig. \ref{FigTSPT}.

\textbf{Discussion} - The current study explores multistability generated by
the modified master equation given by Eq. (\ref{MME}). The proposed modified
master equation (\ref{MME}) can be constructed for any physical system whose
Hilbert space has finite dimensionality. Any candidate master equation
modification has to satisfy some legitimizing properties. For the master
equation given by Eq. (\ref{MME}), the condition $\mathrm{d}\operatorname{Tr}%
\rho/\mathrm{d}t=0$ holds provided that $\operatorname{Tr}\rho=1$ (i.e. $\rho$
is normalized), and $\mathrm{d}\operatorname{Tr}\rho^{2}/\mathrm{d}t=0$,
provided that $\rho^{2}=\rho$ (i.e. $\rho$ represents a pure state). The first
property guaranties norm conservation, whereas positive-semi-definiteness of
$\rho$ is ensured by the second property, together with the relation
$\mathrm{d\log}\left(  \det\rho\right)  /\mathrm{d}t=-2\operatorname{Tr}%
\left(  \Theta-\left\langle \Theta\right\rangle \right)  $ [see Eq. (2.196) of
Ref. \cite{Buks_QMLN}].

Suppression of entanglement (i.e. disentanglement) is introduced using the
operator $\mathcal{Q}^{\left(  \mathrm{D}\right)  }$ \cite{Buks_2400036}. The
added disentanglement term makes the collapse postulate of QM redundant. This
nonlinear term has no effect on any product (i.e. disentangled) state. For a
multipartite system, disentanglement between any pair of subsystems can be
introduced. The expectation value $\langle{\mathcal{Q}^{\left(  \mathrm{D}
\right)  } \rangle}$ is invariant under any subsystem unitary transformation.
The disentanglement operator $\mathcal{Q}^{\left(  \mathrm{D}\right)  }$ can
be constructed for both distinguishable \cite{Buks_2400036} and
indistinguishable particles \cite{Buks_SDIP_up}. Moreover, thermalization can
be incorporated using the operator $\mathcal{Q}^{\left(  \mathrm{H}\right)  }$.

\textbf{Summary} - The current study is motivated by an apparent discrepancy
between some experimental observations and the standard theory of QM.
Multistability has been experimentally observed in a variety of quantum
systems. On the other hand, in standard QM the time evolution is governed by a
monostable ME. For some cases, is has remained unclear how multistability can
be theoretically derived from standard QM.

The spontaneous disentanglement hypothesis is inherently falsifiable, because
it yields predictions, which are experimentally distinguishable from
predictions obtained from standard QM. It is found that multistability can be
obtained in the presence of spontaneous disentanglement. In particular, the
modified ME (\ref{MME}) yields a PT for the TIM (see Figs. \ref{FigTSPT}%
-\ref{FigFSBS}), and a DI for the longitudinally driven spins (see Fig.
\ref{FigTSP}). These theoretical findings, together with experimental
observations of multistability in finite quantum systems, indirectly support
the hypothesis that spontaneous disentanglement occurs in quantum systems.

Useful discussions with Michael Reznikov are acknowledged.

\bibliographystyle{ieeepes}
\bibliography{Eyal_Bib}

\begin{thebibliography}{10}

\bibitem{Fernengel_385701}
Bernd Fernengel and Barbara Drossel,
\newblock ``Bifurcations and chaos in nonlinear lindblad equations'',
\newblock {\em Journal of Physics A: Mathematical and Theoretical}, vol. 53,
  no. 38, pp. 385701, 2020.

\bibitem{Lindblad_119}
Goran Lindblad,
\newblock ``On the generators of quantum dynamical semigroups'',
\newblock {\em Communications in Mathematical Physics}, vol. 48, no. 2, pp.
  119--130, 1976.

\bibitem{Manzano_025106}
Daniel Manzano,
\newblock ``A short introduction to the lindblad master equation'',
\newblock {\em Aip Advances}, vol. 10, no. 2, pp. 025106, 2020.

\bibitem{Grabert_161}
H~Grabert,
\newblock ``Nonlinear relaxation and fluctuations of damped quantum systems'',
\newblock {\em Zeitschrift f{\"u}r Physik B Condensed Matter}, vol. 49, no. 2,
  pp. 161--172, 1982.

\bibitem{Ottinger_052119}
Hans~Christian {\"O}ttinger,
\newblock ``Nonlinear thermodynamic quantum master equation: Properties and
  examples'',
\newblock {\em Physical Review A}, vol. 82, no. 5, pp. 052119, 2010.

\bibitem{Chomaz_68}
Philippe Chomaz and Francesca Gulminelli,
\newblock ``Phase transitions in finite systems'',
\newblock in {\em Dynamics and thermodynamics of systems with long-range
  interactions}, pp. 68--129. Springer, 2002.

\bibitem{mainwood2005phase}
Paul Mainwood,
\newblock ``Phase transitions in finite systems'',
\newblock 2005.

\bibitem{Callender_539}
Craig Callender,
\newblock ``Taking thermodynamics too seriously'',
\newblock {\em Studies in history and philosophy of science part B: studies in
  history and philosophy of modern physics}, vol. 32, no. 4, pp. 539--553,
  2001.

\bibitem{Liu_S92}
Chuang Liu,
\newblock ``Explaining the emergence of cooperative phenomena'',
\newblock {\em Philosophy of Science}, vol. 66, no. S3, pp. S92--S106, 1999.

\bibitem{Ardourel_99}
Vincent Ardourel and Sorin Bangu,
\newblock ``Finite-size scaling theory: Quantitative and qualitative approaches
  to critical phenomena'',
\newblock {\em Studies in History and Philosophy of Science}, vol. 100, pp.
  99--106, 2023.

\bibitem{Shech_1170}
Elay Shech,
\newblock ``What is the paradox of phase transitions?'',
\newblock {\em Philosophy of Science}, vol. 80, no. 5, pp. 1170--1181, 2013.

\bibitem{toda1978statistical_I}
Morikazu Toda, Ryogo Kubo, Nobuhiko Sait{\=o}, Natsuki Hashitsume, and Natsuki
  Hashitsume,
\newblock {\em Statistical physics I},
\newblock Springer Science, 1978.

\bibitem{Roch_633}
Nicolas Roch, Serge Florens, Vincent Bouchiat, Wolfgang Wernsdorfer, and Franck
  Balestro,
\newblock ``Quantum phase transition in a single-molecule quantum dot'',
\newblock {\em Nature}, vol. 453, no. 7195, pp. 633--637, 2008.

\bibitem{Thomas_145}
L~Thomas, FL~Lionti, R~Ballou, Dante Gatteschi, Roberta Sessoli, and B~Barbara,
\newblock ``Macroscopic quantum tunnelling of magnetization in a single crystal
  of nanomagnets'',
\newblock {\em Nature}, vol. 383, no. 6596, pp. 145--147, 1996.

\bibitem{Trishin_236801}
Sergey Trishin, Christian Lotze, Nils Bogdanoff, Felix von Oppen, and
  Katharina~J Franke,
\newblock ``Moir{\'e} tuning of spin excitations: Individual fe atoms on mos
  2/au (111)'',
\newblock {\em Physical Review Letters}, vol. 127, no. 23, pp. 236801, 2021.

\bibitem{Blesio_045113}
GG~Blesio and AA~Aligia,
\newblock ``Topological quantum phase transition in individual fe atoms on mos
  2/au (111)'',
\newblock {\em Physical Review B}, vol. 108, no. 4, pp. 045113, 2023.

\bibitem{Levi_053516}
Roei Levi, Sergei Masis, and Eyal Buks,
\newblock ``Instability in the hartmann-hahn double resonance'',
\newblock {\em Phys. Rev. A}, vol. 102, pp. 053516, Nov 2020.

\bibitem{gurevich2020magnetization}
Alexander~G Gurevich and Gennadii~A Melkov,
\newblock {\em Magnetization oscillations and waves},
\newblock CRC press, 2020.

\bibitem{Suhl_209}
H~Suhl,
\newblock ``The theory of ferromagnetic resonance at high signal powers'',
\newblock {\em Journal of Physics and Chemistry of Solids}, vol. 1, no. 4, pp.
  209--227, 1957.

\bibitem{Schlomann_672}
E~Schl{\"o}mann, RI~Joseph, and I~Bady,
\newblock ``Spin-wave instability in hexagonal ferrites with a preferential
  plane'',
\newblock {\em Journal of Applied Physics}, vol. 34, no. 3, pp. 672--681, 1963.

\bibitem{Zvyagin_174408}
AA~Zvyagin,
\newblock ``Modulation of the longitudinal pumping in quantum spin systems'',
\newblock {\em Physical Review B}, vol. 101, no. 17, pp. 174408, 2020.

\bibitem{Schlomann_S386}
E~Schl{\"o}mann, JJ~Green, and uU~Milano,
\newblock ``Recent developments in ferromagnetic resonance at high power
  levels'',
\newblock {\em Journal of Applied Physics}, vol. 31, no. 5, pp. S386--S395,
  1960.

\bibitem{breuer2002theory}
Heinz-Peter Breuer, Francesco Petruccione, et~al.,
\newblock {\em The theory of open quantum systems},
\newblock Oxford University Press on Demand, 2002.

\bibitem{Drossel_217}
Barbara Drossel,
\newblock ``What condensed matter physics and statistical physics teach us
  about the limits of unitary time evolution'',
\newblock {\em Quantum Studies: Mathematics and Foundations}, vol. 7, no. 2,
  pp. 217--231, 2020.

\bibitem{Hicke_024401}
C~Hicke and MI~Dykman,
\newblock ``Classical dynamics of resonantly modulated large-spin systems'',
\newblock {\em Physical Review B}, vol. 78, no. 2, pp. 024401, 2008.

\bibitem{Klobus_034201}
Waldemar K{\l}obus, Pawe{\l} Kurzy{\'n}ski, Marek Ku{\'s}, Wies{\l}aw
  Laskowski, Robert Przybycie{\'n}, and Karol {\.Z}yczkowski,
\newblock ``Transition from order to chaos in reduced quantum dynamics'',
\newblock {\em Physical Review E}, vol. 105, no. 3, pp. 034201, 2022.

\bibitem{Hush_061401}
Michael~R Hush, Weibin Li, Sam Genway, Igor Lesanovsky, and Andrew~D Armour,
\newblock ``Spin correlations as a probe of quantum synchronization in
  trapped-ion phonon lasers'',
\newblock {\em Physical Review A}, vol. 91, no. 6, 2015.

\bibitem{Vedral_22}
Vlatko Vedral,
\newblock ``Mean-field approximations and multipartite thermal correlations'',
\newblock {\em New Journal of physics}, vol. 6, no. 1, pp. 22, 2004.

\bibitem{Sakthivadivel_035}
Dalton~AR Sakthivadivel,
\newblock ``Magnetisation and mean field theory in the ising model'',
\newblock {\em SciPost Physics Lecture Notes}, p. 035, 2022.

\bibitem{Osacar_045404}
C~Os{\'a}car and AF~Pacheco,
\newblock ``A mean field approach to the ising chain in a transverse magnetic
  field'',
\newblock {\em European Journal of Physics}, vol. 38, no. 4, pp. 045404, 2017.

\bibitem{Buks_2400036}
Eyal Buks,
\newblock ``Spontaneous disentanglement and thermalization'',
\newblock {\em Advanced Quantum Technologies}, p. 2400036, 2024.

\bibitem{Kaplan_055002}
David~E Kaplan and Surjeet Rajendran,
\newblock ``Causal framework for nonlinear quantum mechanics'',
\newblock {\em Physical Review D}, vol. 105, no. 5, pp. 055002, 2022.

\bibitem{Geller_2200156}
Michael~R Geller,
\newblock ``Fast quantum state discrimination with nonlinear positive
  trace-preserving channels'',
\newblock {\em Advanced Quantum Technologies}, p. 2200156, 2023.

\bibitem{Grimaudo_033835}
R~Grimaudo, Asm De~Castro, M~Ku{\'s}, and A~Messina,
\newblock ``Exactly solvable time-dependent pseudo-hermitian su (1, 1)
  hamiltonian models'',
\newblock {\em Physical Review A}, vol. 98, no. 3, pp. 033835, 2018.

\bibitem{Kowalski_167955}
K~Kowalski and J~Rembieli{\'n}ski,
\newblock ``Integrable nonlinear evolution of the qubit'',
\newblock {\em Annals of Physics}, vol. 411, pp. 167955, 2019.

\bibitem{Jaynes_579}
Edwin~T Jaynes,
\newblock ``The minimum entropy production principle'',
\newblock {\em Annual Review of Physical Chemistry}, vol. 31, no. 1, pp.
  579--601, 1980.

\bibitem{Buks_052217}
Eyal Buks and Dvir Schwartz,
\newblock ``Stability of the grabert master equation'',
\newblock {\em Physical Review A}, vol. 103, no. 5, pp. 052217, 2021.

\bibitem{Lieb_407}
Elliott Lieb, Theodore Schultz, and Daniel Mattis,
\newblock ``Two soluble models of an antiferromagnetic chain'',
\newblock {\em Annals of Physics}, vol. 16, no. 3, pp. 407--466, 1961.

\bibitem{Pfeuty_79}
Pierre Pfeuty,
\newblock ``The one-dimensional ising model with a transverse field'',
\newblock {\em ANNALS of Physics}, vol. 57, no. 1, pp. 79--90, 1970.

\bibitem{Mbeng_2009_09208}
Glen~Bigan Mbeng, Angelo Russomanno, and Giuseppe~E Santoro,
\newblock ``The quantum ising chain for beginners'',
\newblock {\em arXiv preprint arXiv:2009.09208}, 2020.

\bibitem{Nagaj_214431}
Daniel Nagaj, Edward Farhi, Jeffrey Goldstone, Peter Shor, and Igor Sylvester,
\newblock ``Quantum transverse-field ising model on an infinite tree from
  matrix product states'',
\newblock {\em Physical Review B}, vol. 77, no. 21, pp. 214431, 2008.

\bibitem{chakrabarti2008quantum}
Bikas~K Chakrabarti, Amit Dutta, and Parongama Sen,
\newblock {\em Quantum Ising phases and transitions in transverse Ising
  models}, vol.~41,
\newblock Springer Science \& Business Media, 2008.

\bibitem{Grimmett_305}
Geoffrey~R Grimmett, Tobias~J Osborne, and Petra~F Scudo,
\newblock ``Entanglement in the quantum ising model'',
\newblock {\em Journal of Statistical Physics}, vol. 131, pp. 305--339, 2008.

\bibitem{Jian_134206}
Chao-Ming Jian, Bela Bauer, Anna Keselman, and Andreas~WW Ludwig,
\newblock ``Criticality and entanglement in nonunitary quantum circuits and
  tensor networks of noninteracting fermions'',
\newblock {\em Physical Review B}, vol. 106, no. 13, pp. 134206, 2022.

\bibitem{Latorre_0304098}
Jos{\'e}~Ignacio Latorre, Enrique Rico, and Guifr{\'e} Vidal,
\newblock ``Ground state entanglement in quantum spin chains'',
\newblock {\em arXiv preprint quant-ph/0304098}, 2003.

\bibitem{Osborne_032110}
Tobias~J Osborne and Michael~A Nielsen,
\newblock ``Entanglement in a simple quantum phase transition'',
\newblock {\em Physical Review A}, vol. 66, no. 3, pp. 032110, 2002.

\bibitem{Parez_2402_06677}
Gilles Parez and William Witczak-Krempa,
\newblock ``The fate of entanglement'',
\newblock {\em arXiv preprint arXiv:2402.06677}, 2024.

\bibitem{kittel1976introduction}
Charles Kittel et~al.,
\newblock {\em Introduction to solid state physics}, vol.~8,
\newblock Wiley New York, 1976.

\bibitem{Kitzinger_033504}
Jonas Kitzinger, Manish Chaudhary, Manikandan Kondappan, Valentin Ivannikov,
  and Tim Byrnes,
\newblock ``Two-axis two-spin squeezed states'',
\newblock {\em Physical Review Research}, vol. 2, no. 3, pp. 033504, 2020.

\bibitem{Kitagawa_5138}
Masahiro Kitagawa and Masahito Ueda,
\newblock ``Squeezed spin states'',
\newblock {\em Physical Review A}, vol. 47, no. 6, pp. 5138, 1993.

\bibitem{Ma_89}
Jian Ma, Xiaoguang Wang, Chang-Pu Sun, and Franco Nori,
\newblock ``Quantum spin squeezing'',
\newblock {\em Physics Reports}, vol. 509, no. 2-3, pp. 89--165, 2011.

\bibitem{Buks_QMLN}
Eyal Buks,
\newblock {\em Quantum mechanics - Lecture Notes},
\newblock http://buks.net.technion.ac.il/teaching/, 2024.

\bibitem{Buks_SDIP_up}
Eyal Buks,
\newblock ``Disentanglementspontaneous disentanglement of indistinguishable
  particles'',
\newblock {\em unpublished}, 2024.

\end{thebibliography}

\end{document}